\documentclass[%
 aip,
rsi,%
 amsmath,amssymb,
 reprint,%
]{revtex4-1}
\usepackage{siunitx}
\usepackage{graphicx}
\usepackage{dcolumn}
\usepackage{bm}
\begin{document}

\preprint{APS/123-QED}

\title{Spatiotemporal plasma hologram}
\author{Zhaohui Wu}
\affiliation{National key laboratory of plasma physics, Research Center of Laser Fusion, China Academy of Engineering Physics, Mianyang, Sichuan,621900,China}
\author{Hao Peng}
\affiliation{Shenzhen Key Laboratory of Ultraintense Laser and Advanced Material Technology,
Center for Advanced Material Diagnostic Technology, and College of Engineering Physics,
Shenzhen Technology University,Shenzhen 518118, China}
\author{Xiaoming Zeng}
\author{Zhaoli Li}
\author{Xiaodong Wang}
\author{Xiao Wang}%
\author{Jie Mu}
\author{Yanlei Zuo}%
\author{Kainan Zhou}%
\affiliation{National key laboratory of plasma physics, Research Center of Laser Fusion, China Academy of Engineering Physics, Mianyang, Sichuan, 621900, China}
\author{Nathaniel J. Fisch}%
\affiliation{Department of Astrophysical Sciences, Princeton University, Princeton, New Jersey 08540}
\author{C. Riconda}
\affiliation{LULI, Sorbonne Universit¨¦, CNRS, ¨¦cole Polytechnique, CEA, F-75005, Paris, France}
\author{S. Weber}
\affiliation{ELI Beamlines facility, Extreme Light Infrastructure ERIC, 25241 Dolni Brezany, Czech Republic}

\date{\today}
\begin{abstract}
We present the first experimental realization of a four-dimensional (4D) plasma hologram capable of recording and reconstructing the full spatiotemporal information of intense laser pulses. The holographic encoding is achieved through the interference of a long object pulse and a counterpropagating short reference pulse, generating an ionized plasma grating that captures both spatial and temporal characteristics of the laser field. A first-order diffractive probe enables the retrieval of encoded information, successfully reconstructing the spatiotemporal profiles of Gaussian and Laguerre-Gaussian beams. The experiment demonstrates the ability to encode artificial information into the laser pulse via spectral modulation and retrieve it through plasma grating diffraction, highlighting potential applications in ultraintense optical data processing. Key innovations include a single-shot, background-free method for direct far-field(focus) spatiotemporal measurement and the observation of laser focus propagation dynamics in plasma. The plasma grating exhibits a stable lifetime of 30-40 ps and supports high repetition rates, suggesting usage for high-speed optical switches and plasmatic analog memory. These advancements establish plasma holography as an advanced platform for ultrafast laser manipulation, with implications for secure optical communication, analog computing, and precision spatiotemporal control of high-intensity lasers.
\end{abstract}
\maketitle
The development of ultra-high-intensity lasers has led to significant challenges for optical components, primarily due to material damage. To mitigate this issue, both optical components and laser facilities have become increasingly large, resulting in elevated manufacturing costs and a range of undesirable effects, including high-order dispersion, surface distortion, chromatic aberration, and spectral phase distortion. Plasma optics offers a promising solution to these challenges due to its robustness in the presence of extremely strong optical fields. Specifically, plasma optics can withstand laser intensities several orders of magnitude higher than traditional solid optics, making them increasingly valuable in applications involving intense and compact laser manipulation. These include, but are not limited to, contrast improvement \cite{Edwards24}, laser compression \cite{Zhaohui22, Andreev06, Malkin991, JUN07, Trines111,Toroker14,Matthew15, Marques19, Wu24PRR}, polarization control \cite{Turnbull16, Turnbull17, Lehmann18}, energy exchange \cite{Liu10}, particle acceleration, tight focusing \cite{Henri19, Edwards22}, and quantum entanglement\cite{qu2024entangled}.

 Plasma optics also enables the storage and retrieval of ultra-strong laser fields,  known as plasma holography. Recent advancements include the demonstration of a three-dimensional (3D) hologram using surface plasma gratings, highlighting the potential of plasmas for transient information storage \cite{Leblanc17}. However, surface plasma  hologram still can not record the temporal characteristics of the incident light. Additionally, the vulnerability of solid surfaces to damage renders this approach less suitable for high-repetition-rate experiments. To address these limitations, gas-based volume plasma gratings have been proposed as a means to achieve four-dimensional (4D) holography. This technique involves imprinting the complete light field information onto a plasma wave, utilizing backscattering in the Raman or Brillouin regimes \cite{Dodin02prl,dodin2002dynamic,Lehmann19}. Previous theoretical studies suggest that this method could enable 4D plasma holography. However, this concept has yet to be experimentally demonstrated due to the challenges associated with controlling the plasma wave.

\begin{figure}
\centering
\includegraphics[width=3.4 in]{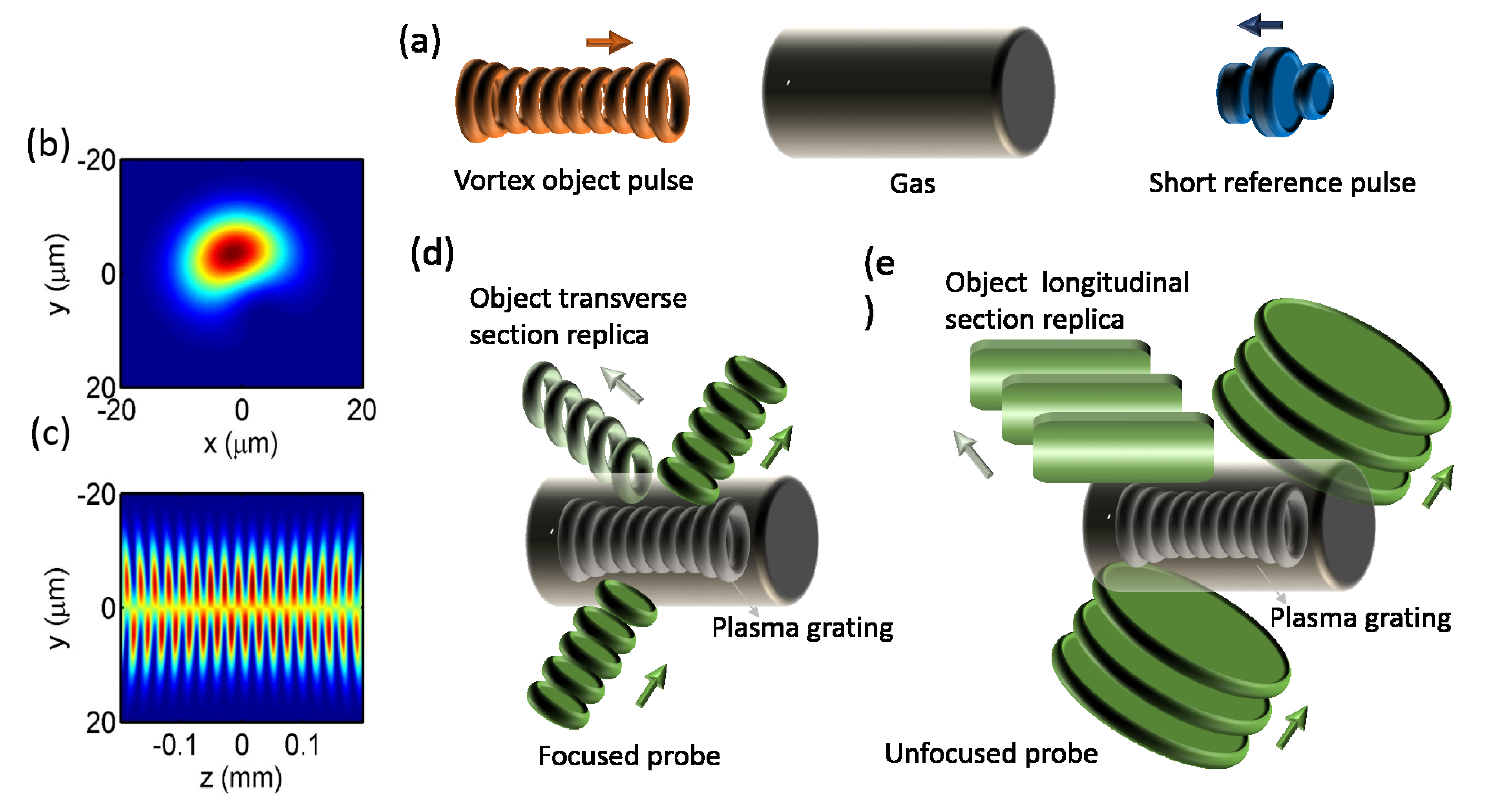}
\caption{(a)Plasma grating created by interfering ionization of a long object pulse and a short reference pulse.(b)-(c), the transverse and longitudinal sections of the interference field with a Laguerre-Gaussian object focus and a uniform interference focus. (d)Reconstruction of the object transverse section with a focused probe. (e)Reconstruction of the object longitudinal section with a unfocused probe.
\label{fig:scheme}}
\end{figure}

 In recent years, a novel type of plasma grating has been developed based on spatially varying ionization\cite{Liping11, Chaojie21, Edwards23, Edwards24}. This plasma grating exhibits exceptional characteristics, including high diffractive efficiency and excellent beam quality, due to the reduced nonlinear effects and instabilities in its generation process. In this work, we propose the use of such plasma gratings for volume plasma holography. The proposed scheme is illustrated in Fig.\ref{fig:scheme}. Initially, an object pulse and a shorter counterpropagating reference pulse, both with the same central wavelength, are focused into the background gas, as depicted in Fig.\ref{fig:scheme}(a). The standing-wave of them forms a volume interference field, the transverse and longitudinal sections of which are shown in Fig.\ref{fig:scheme}(b) and Fig.\ref{fig:scheme}(c), respectively. As plasma density is proportional to the interference intensity, the plasm grating would record the light field during the ionization process. After several picoseconds, a probe pulse is injected into the plasma grating at the Bragg angle. The zero-order light will pass through the plasma grating unchanged, while the first-order light, reflected by the plasma grating, will carry the light information. Since the diffractive efficiency is proportional to both the modulation depth and the interference field, the light information can be shown by the intensity of the diffractive signal. The cross-sectional and longitudinal profiles of the laser pulses can be retrieved by the focusing or collimated probes, as shown in Fig. \ref{fig:scheme}(c) and Fig. \ref{fig:scheme}(d), respectively.

The relationship between the diffracted probe and the object pulse can be derived from the diffraction efficiency of the plasma grating. For an ionized plasma grating, the interference intensity formed by the object and reference pulses can be expressed as $\sqrt{I_0 I_1} \cos\left(2\pi z/\Lambda\right)$, where $\Lambda$ is the spatial period of the interference pattern, and $I_0$ and $I_1$ are the intensities of the object and reference pulses, respectively. The diffraction efficiency of the reflective Bragg grating is given by $\eta \simeq n_1^2 \pi^2 L^2 / (\lambda^2 \cos^2 \theta_B)$, where $\lambda$ is the wavelength of the probe, $\theta_B$ is the Bragg angle, $L$ is the thickness of the grating, and $n_1$ is the effective refractive index modulation of the plasma grating \cite{Edwards23,clark02,Clark032}. It can be obtained by expressing the refractive index $n$ as a sum of Fourier components with spatial periods $m\Lambda$ ($m = 0, \pm1, \pm2, \ldots$). Given that $n = \sqrt{1 - n_e / n_c} \simeq n_e / 2n_c=\rho P/2n_c$ when $n_e \ll n_c$, where $n_e$ is the electron density, $n_c = \varepsilon_0 m_e c^2 / e^2 \lambda^2$ is the critical density for the probe, $m_e$ and $e$ is the electron mass and charge, $\varepsilon_0$ is the vacuum permittivity, $c$ is the light velocity in vacuum, $\lambda$ is probe wavelength, $\rho$ is the gas density and $P$ is the ionization probability per molecule. For laser-ionized plasma, $P$ can be well evaluated using the Perelomov-Popov-Terent\'{e}v(MO-PPT) model\cite{zhao2014multiphoton,zhao2016analytical,zhao2016kinetics,popov2004tunnel,tong2005empirical,zhou2009molecular}, which describes a complex dependence on $I_0$. However, simulation results show that the first-order Fourier component $P$ exhibits an approximately linear response to $\sqrt{I_0}$ for laser intensities exceeding $10^{14}~\rm{W/cm^2}$(see the supplement materia).
 Consequently, the $\eta$ can be approximately expressed as follows:
\begin{eqnarray}
\begin{array}{rcl}
\eta\propto I_0 I_1\rho^2\pi^2L^2/(\lambda^2 cos^2\theta_B).
\end{array}
\label{diffaction1}
\end{eqnarray}

If the reference pulse and probe are sufficiently uniform, it has $I\propto I_0$. In the case that the object pulse has a spatial phase $\phi(x,y)$, both the interfering field and the plasma grating will bench to encode the phase information, as shown in Fig.\ref{fig:scheme}(c). This information can also reconstructed at the wavefront of the diffracted light with $I\propto I_0 e^{i\phi(x,y)}$.

\begin{figure}[htpb]
\centering
\includegraphics[width=3 in]{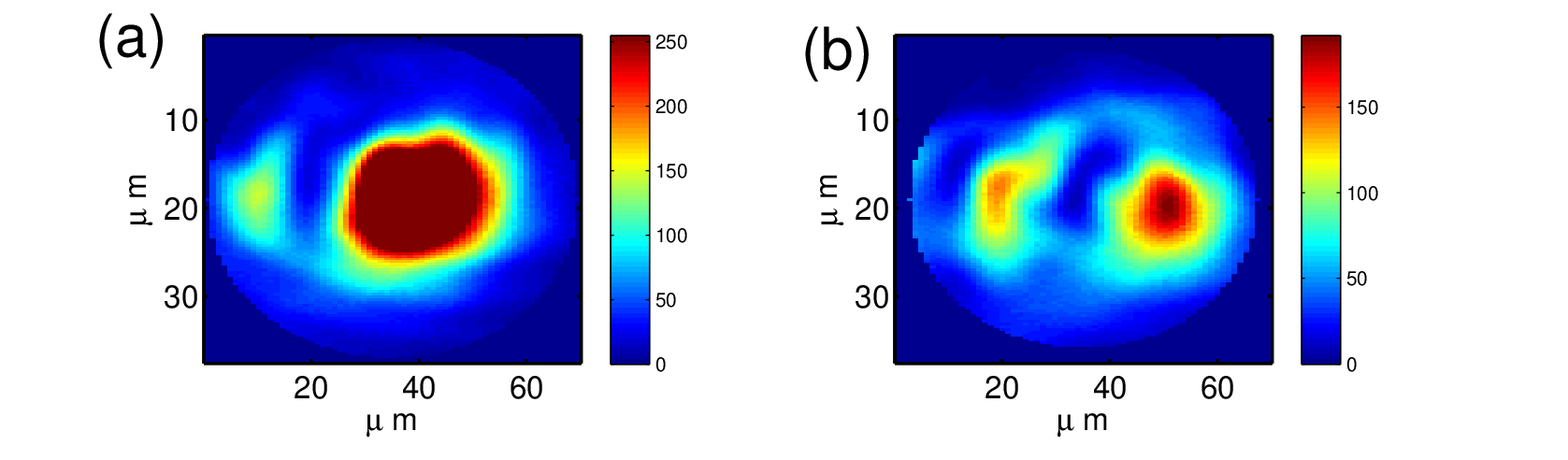}
\includegraphics[width=3 in]{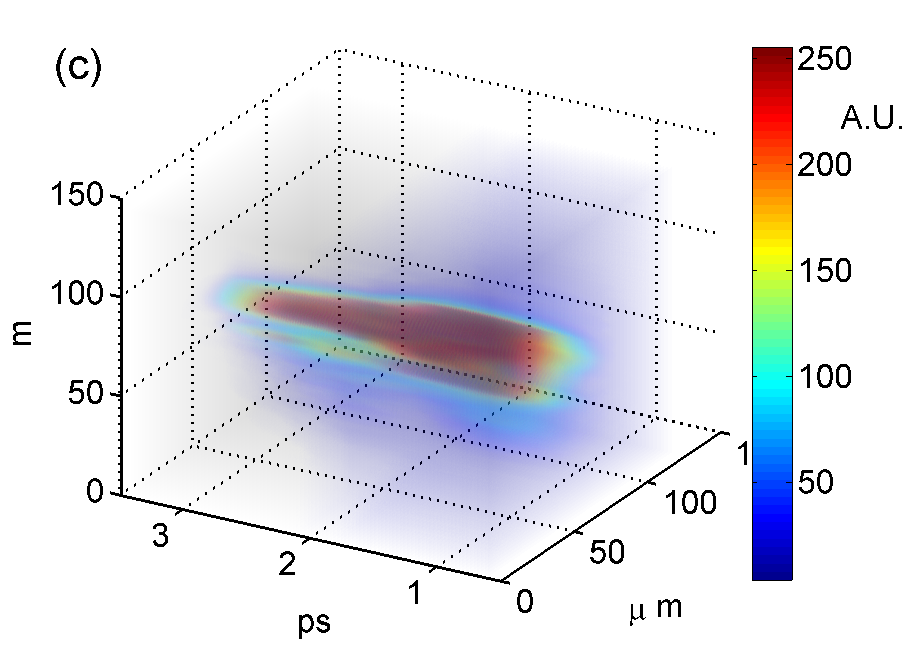}\\
\includegraphics[width=3 in]{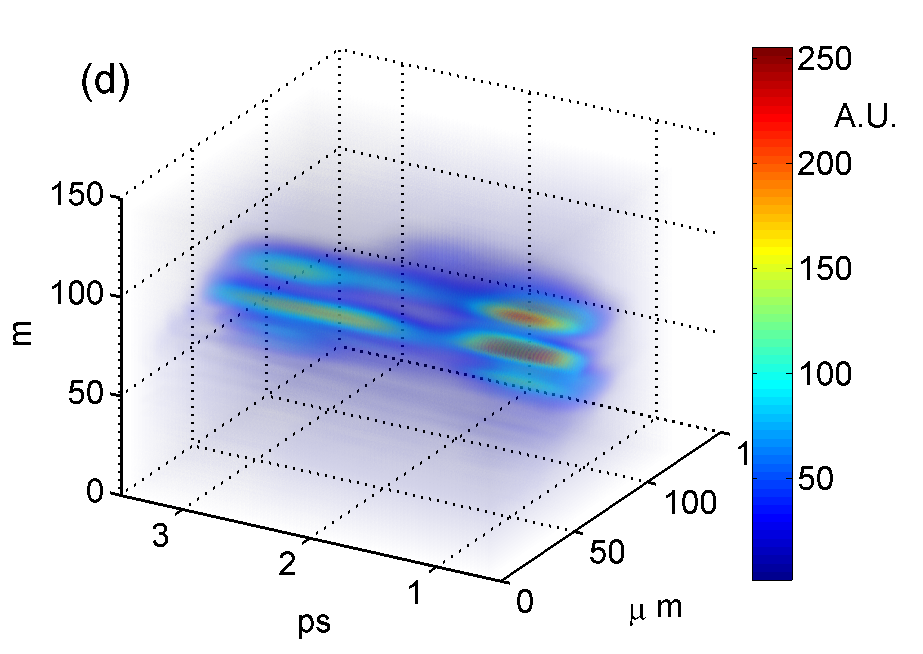}\\
\caption{(a) and (b), retrieved beam waists of a Gaussian focus and a  Laguerre-Gaussian focuses. (c) and (d), 3D spatiotemporal structure of a Gaussian and a Laguerre-Gaussian focuses.
\label{fig:3Dfield}}
\end{figure}

 The proposed scheme was experimentally demonstrated using a Ti:sapphire CPA laser facility(the experiment setup is shown in the supplement materia). Both the object pulse and reference pulse originated from the same source, with a central wavelength of 800 nm and a bandwidth of 80 nm. The laser output was equally split into two pulses: one pulse, with an energy up to 6 mJ, was directly focused into the air using an f/15 lens for the object pulse. It had a FWHM focus diameter of $\sim20~\rm{\mu m}$ and a full width at half maximum (FWHM) duration of several picoseconds. The peak laser focused intensity is $(3-4)\times10^{14}~\rm{W/cm^2}$ by adjusted the pulse energy and duration.  The other pulse, used as the reference pulse, was first compressed to approximately 30 fs and then cut off to a diameter of $1~\rm{mm}$ by an aperture. As a result, it had an energy of $60~\rm{\mu J}$. It was then focused to the same region using an f/100 lens, corresponding to a FWHM focused diameter $\sim100~\rm{\mu m}$, and the focused intensity $2.5\times10^{13}~\rm{W/cm^2}$. Due to the reference pulse's focus size being five times larger than that of the object pulse, the beam was nearly uniform within the interference region, ensuring that the interference field only recorded information from the object pulse. Both pulses had an adjustable delay line to precisely synchronize their timing.The probe pulse was generated from the short reference pulse. It was then doubled in frequency using a BBO crystal, resulting in a central wavelength of 420 nm and an FWHM spectral width of 10 nm. This probe pulse was directed toward the plasma grating at a Bragg angle of approximately $60^{\circ}$. The intensity is kept below $10^{13}~\rm{W/cm^2}$ in case of influencing the plasma grating. The first-order diffracted pulse was then collected using a lens and delivered to a CCD camera for further analysis and imaging.

Although the diffracted probe carries the full 3D light field information, only the 2D profile can be captured by the CCD camera. To extract the full light field information, we first focused the probe to a specific cross-section of the plasma grating. The probe, focused with an f/40 lens, had a beam waist slightly larger than the cross-section of the plasma grating. As a result, the first-order diffracted probe was able to deliver the information from this specific section to the CCD camera.To confirm this, we employed a spiral phase plate with topological charge $m=1$, creating a Laguerre-Gaussian object pulse with a characteristic ring-shaped structure at the beam waist. The reconstructed object waist successfully replicated the ring structure when the phase plate was used, as shown in Fig. \ref{fig:3Dfield}(a) and Fig. \ref{fig:3Dfield}(b). The ring position of both the initial and retrieved object waists changed as the phase plate was slightly tilted(see the supplement material), demonstrating that the variation of the object focus could be precisely captured. Subsequently, the probe focus was translated along the plasma grating to record each cross-section of the object focus. By combining these individual sections, the full 3D light field was reconstructed. As shown in Fig. \ref{fig:3Dfield}(a) and Fig. \ref{fig:3Dfield}(b), the reconstructed laser focus revealed the temporal-spatial structure of the focuses for both a normal Gaussian beam and a Laguerre-Gaussian beam. To the best of our knowledge, this represents the first experimental direct observation of the spatiotemporal structure of a laser focus.

\begin{figure}[htpb]
\centering
\includegraphics[width=3 in]{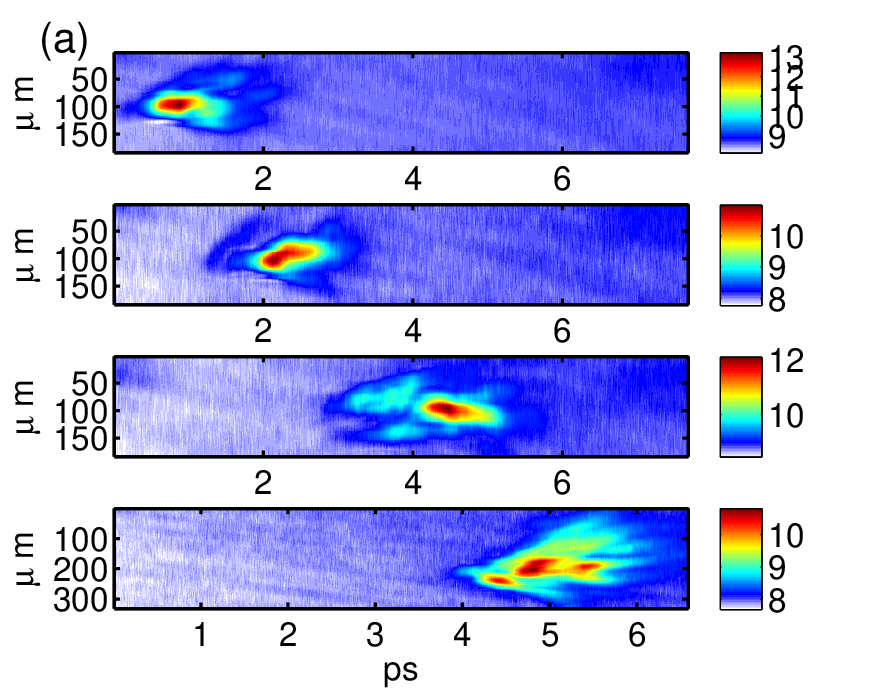}\\
\includegraphics[width=3 in]{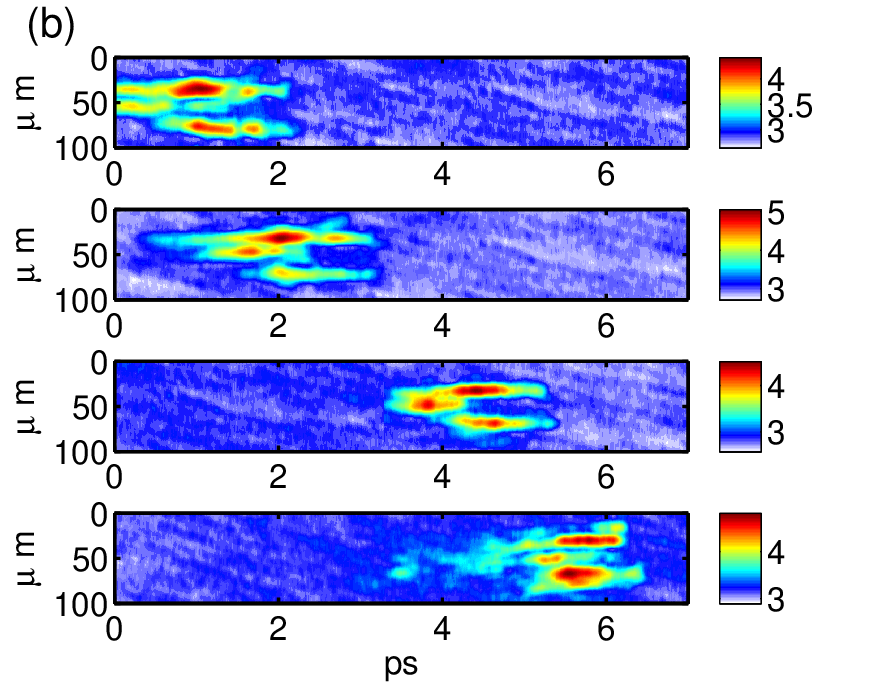}\\
\caption{Propagation process of a Gaussian object focus(a) and a Laguerre-Gaussian focuses(b) in plasma.
\label{fig:2Dfield}}
\end{figure}

To detect the spatiotemporal structure of the laser focus in a single-shot measurement, we used a pinhole to create a uniform and large probe beam, which was able to cover the entire plasma grating. Since the probe was incident at a $60^\circ$ angle relative to the plasma grating, it allowed the temporal and spatial information of the object pulse to be projected longitudinally and transversely onto the CCD camera, respectively. Additionally, by adjusting the delays between the object and reference pulses, they could meet at different positions in the plasma, which allowed us to capture an ultrafast process: propagation of laser focus, including both Gaussian and Laguerre-Gaussian beams, in the plasma. As shown in Fig. \ref{fig:2Dfield}(a), the Gaussian focus initially focused and then diverged as it propagated through the plasma, which provides insight into the beam's behavior in a nonlinear medium. In contrast to the near-field beam, the far-field laser pulse exhibited strong temporal-spatial coupling, meaning that its temporal shape changed during the focusing process. For the Laguerre-Gaussian object beam in our experiment, as the central circle is asymmetric [see Fig.\ref{fig:3Dfield}(a)], the vortex focus was split into two distinct lobes in the upper and lower regions. Consequently, when the probe beam was incident from the side, the characteristic hollow structure was observed, as shown in Fig. \ref{fig:2Dfield}(b). This hollow structure is a signature feature of the orbital angular momentum carried by the Laguerre-Gaussian mode. Different from the Gaussian focus, the vortex focus has a slightly larger size and can propagate longer without divergence, as shown in Fig. \ref{fig:2Dfield}(b).

\begin{figure}[htpb]
\centering
\includegraphics[width=1.6 in]{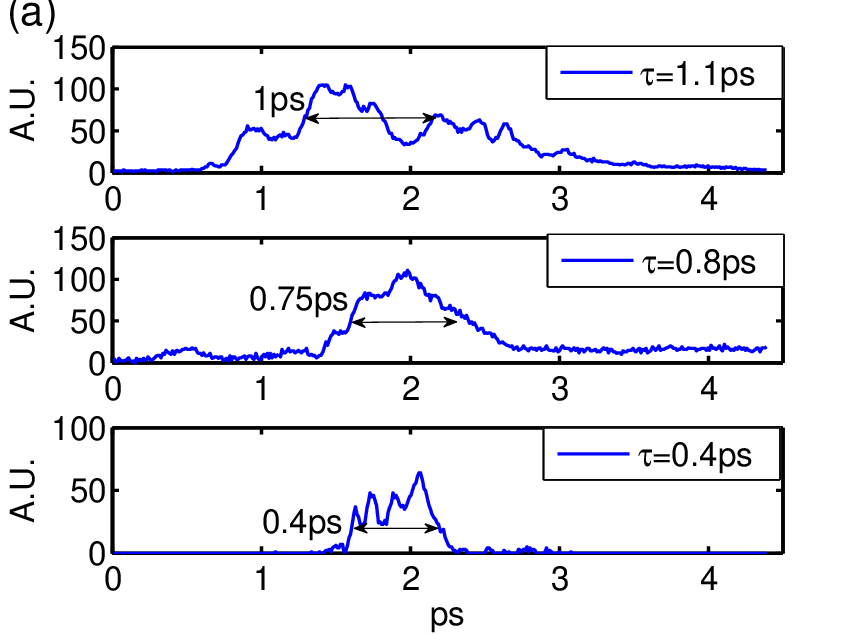}
\includegraphics[width=1.6 in]{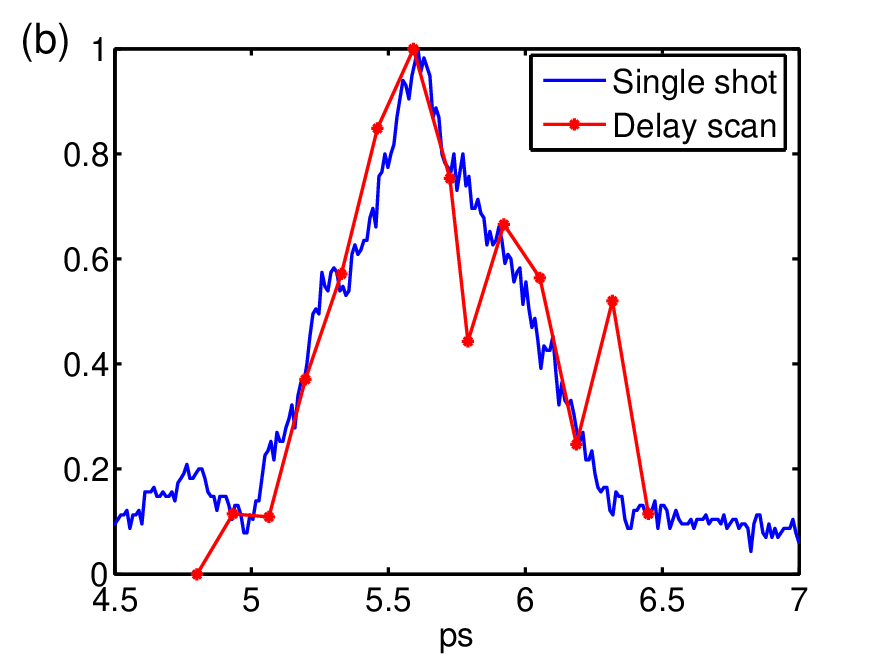}\\
\includegraphics[width=1.6 in]{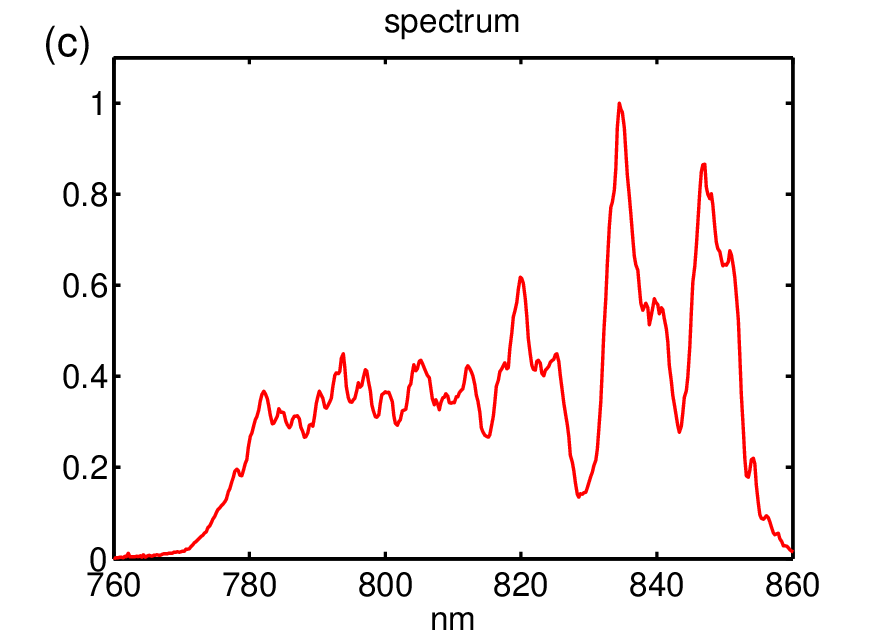}
\includegraphics[width=1.6 in]{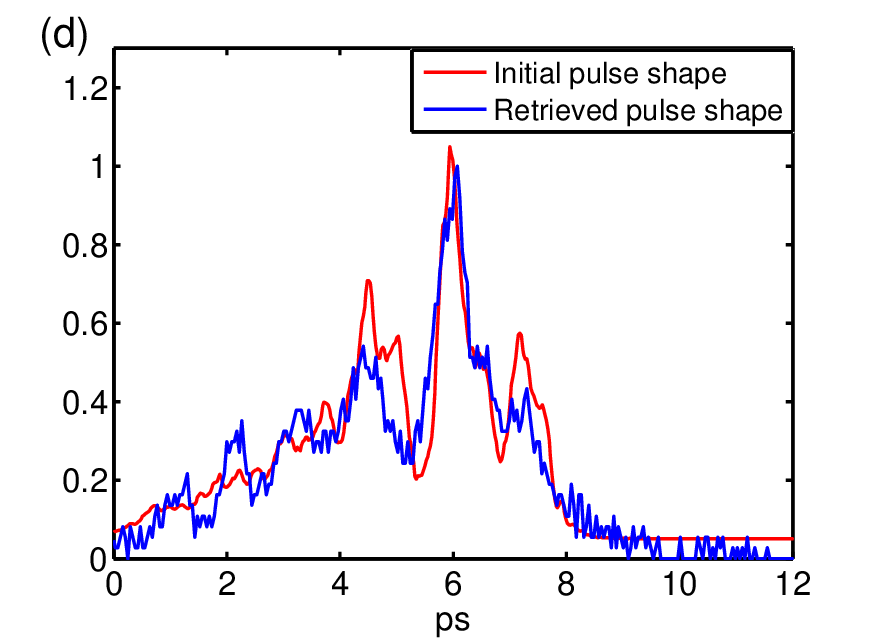}\\
\includegraphics[width=1.6 in]{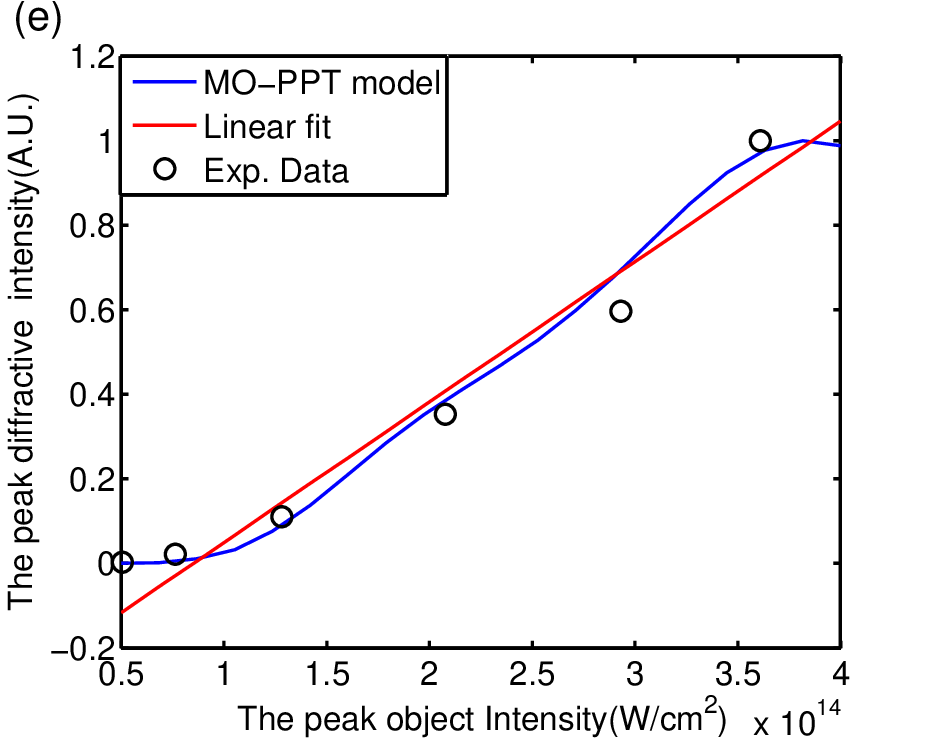}
\includegraphics[width=1.6 in]{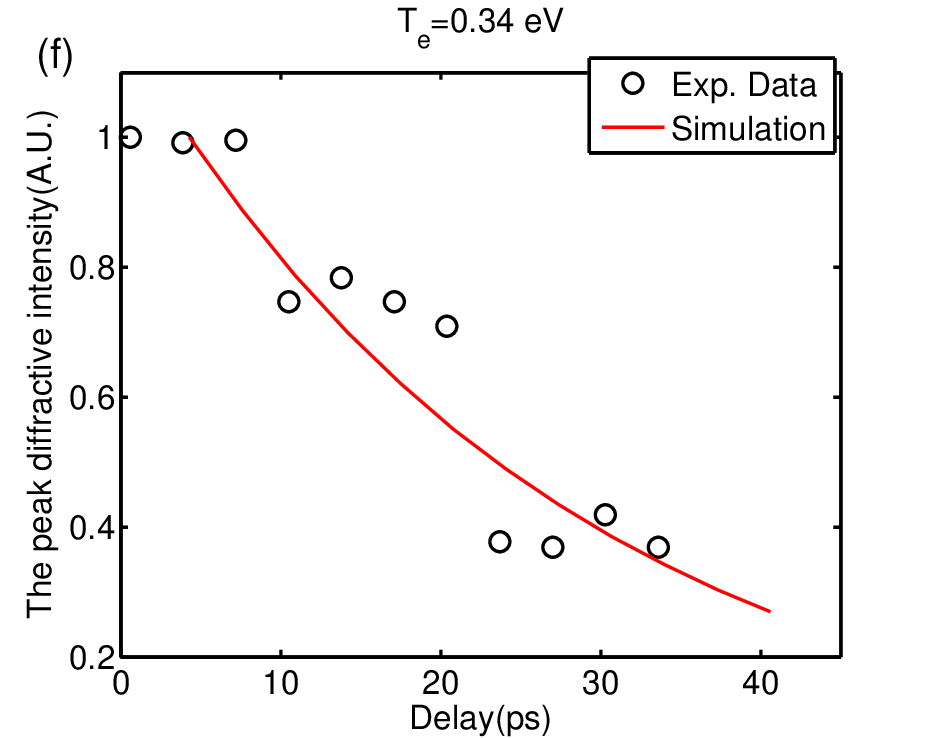}\\
\caption{(a)The retrieved  temporal shape of the far-field object with different durations. (b)Normalized pulse temporal shape of single-shot and delay scanning. (c)The near-field object spectrum.(d)The input and retrieved far-field pulse shape at the beam centerline. (e)Simulation and experimental peak diffractive intensities with increasing object peak intensity. (f)Simulation and experimental peak diffractive intensities with various probe delays.
\label{fig:grating}}
\end{figure}

The correction of the reconstructed temporal pulse shape is proved by two ways. Firstly, we tested it with different pulse durations by adjusting the grating distance in the in the Ti:sapphire laser system. For a chirp pulse, the duration is roughly given as $\tau=kD$, where D is the grating distance from position of the zero pulse chirp, and $k\approx1\rm{ps/mm}$ is the coefficient determined by the laser system. As shown in Fig.\ref{fig:grating}(a), with the input pulse duration of 0.4 ps, 0.8 ps and 1.1 ps, the retrieved pulse durations were 0.4 ps, 0.75 ps and 1ps, indicating the plasma hologram can well return the pulse durations. Moreover, we focused the probe to one point of the plasma grating and adjusted the reference pulse delay. Similar as a scanning second-order autocorrelator, the pulse shape was obtained by using the reference pulse to scan along the object pulse at this point. The result of single shot also matches well with that of the scanning at $\tau=0.75$ ps,  as shown in Fig.\ref{fig:grating}(b).

By employing a vortex focus, volume plasma holography has been previously demonstrated to retrieve spatial far-field light information. Here, we extend this capability to replicate the fine temporal far-field structure of the laser pulse. For a chirped object pulse, if the duration is sufficiently long, the near-field pulse intensity is approximately $\beta {I_0}_\omega$, where ${I_0}_\omega$ represents the spectral intensity and $\beta$ is the pulse chirp. For a nearly Gaussian laser beam, the far-field intensity at the beam center can be expressed as ${I_0}_\omega\beta/\sqrt{1-z^2/z^2_R}$, where $z$ is the propagation distance from the waist and $z_R$ is the Rayleigh length. To verify that the fine temporal structure of the object pulse can be retrieved, we adjusted the laser system's spectrum to form a multi-peak structure, as shown in Fig.\ref{fig:grating}(c). Subsequently, with a pulse duration close to 4 ps, the input and reconstructed far-field pulse shapes were obtained from the spectrum and the plasma hologram, respectively. As illustrated in Fig.\ref{fig:grating}(d), the reconstructed temporal waveform closely matches the original spectral profile, demonstrating that intricate temporal structures can be encoded in the ultraintense laser and subsequently recovered via volume plasma holography.

To calibrate the relationship between the object intensity and the diffracted intensity, the MO-PPT model was employed to simulate the tunneling ionization process in the air. The results show that the diffracted intensity increases approximately linearly with the object intensity in the range of $(0.5-4)\times10^{14}\rm{W/cm^2}$. To verify this, we measured the peak intensity of the diffracted light for object intensities within the same range, $(0.5-4)\times10^{14}\rm{W/cm^2}$, which agrees well with the MO-PPT model and the linear fitting curve, as shown in Fig.\ref{fig:grating}(e). In the experiment, the maximum object intensity reached $4\times10^{14}\rm{W/cm^2}$, which lies between the ionization thresholds of the first and second electrons of nitrogen and oxygen. Ionized plasma gratings may still function at higher intensities once inner-shell electrons are ionized. However, this process is considerably more complex and warrants further investigation. The maximum diffraction efficiency obtained in the experiment was close to $2\%$, and it can be further enhanced by increasing the grating thickness and gas density while reducing the incident angle of the probe, as shown in Eq.~\ref{diffaction1}.

With various probe delay, we can observe the degradation process of the plasma grating. Both the grating modulation and the diffracted signal gradually decreased due to the ambipolar expansion of the plasma, as confirmed by previous studies on ionized plasma gratings \cite{Chaojie21}. This process can be simply modeled as plasma expansion driven by thermal pressure. The decreasing diffractive signal is given by
\begin{eqnarray}
\begin{array}{rcl}
I \propto \delta n^2_e\propto (e^{-2tC_s/\Lambda})^2,
\end{array}
\label{delay}
\end{eqnarray}
where $C_s = \sqrt{T_e/m_i}$ is the ion acoustic velocity, and $m_i$ the ion mass. $T_e$ representing the transverse electron temperature can be estimated by the electron energy driven by ponderomotive force of the interference field. The transverse electron velocity is given as $v=F\tau/m_e=2\pi c^2a_0a_1/\Lambda sin(2\pi z/\Lambda)$, where $F$ is the ponderomotive force, $a_0$ and $a_1$ are the vector potentials of object and
 reference normalized by $m_ec^2/e$, $\tau\approx15 fs$ is the half duration of the reference pulse. The maximum $T_e$ is about 0.34 eV when the object intensity is around $3.5\times10^{14}~\rm{W/cm^2}$. By employing the temperature into Eq.\ref{delay}, the peak diffractive intensity decreasing with the probe delay matches well with the experiment, as shown in Fig.\ref{fig:grating}(f). Due to plasma expansion, the ionized plasma grating in air has a lifetime of 30-40 ps, suggesting its potential use as a flash memory element or an optical switch on this timescale.

In conclusion, we present a novel approach to volume plasma holography, generated by interference-induced ionization, which can imprint, store, and retrieve the full spatiotemporal information of ultra-intense laser pulses. In the experiment, we successfully reconstructed the far-field 3D distributions of both Gaussian and Laguerre-Gaussian beams. Furthermore, we introduced a single-shot, background-free method for directly measuring far-field laser spatiotemporal information, enabling the direct observation of the propagation process of the laser focus in plasma. This technique opens new avenues for ultrafast optical manipulation and precise spatiotemporal control of high-intensity laser pulses.

By employing vortex beam focusing and multi-peak pulse temporal shapes, we demonstrated that artificial information can be encoded into ultra-intense laser pulses and subsequently retrieved using volume plasma holography. By modifying the grating structure, this technique can be employed to reshape laser beams, offering precise control over their spatiotemporal properties. Moreover, it offers a high-repetition-rate, erasable plasmatic analog memory that could be applied in encrypted optical communication technologies or analog computing. This capability marks a significant step toward developing advanced, high-speed optical systems for secure communication and computational tasks.

This work was partly supported by the Science and Technology on Plasma Physics Laboratory.

\bibliography{ref}

\providecommand{\noopsort}[1]{}\providecommand{\singleletter}[1]{#1}%
\begin{thebibliography}{32}%
\makeatletter
\providecommand \@ifxundefined [1]{%
 \@ifx{#1\undefined}
}%
\providecommand \@ifnum [1]{%
 \ifnum #1\expandafter \@firstoftwo
 \else \expandafter \@secondoftwo
 \fi
}%
\providecommand \@ifx [1]{%
 \ifx #1\expandafter \@firstoftwo
 \else \expandafter \@secondoftwo
 \fi
}%
\providecommand \natexlab [1]{#1}%
\providecommand \enquote  [1]{``#1''}%
\providecommand \bibnamefont  [1]{#1}%
\providecommand \bibfnamefont [1]{#1}%
\providecommand \citenamefont [1]{#1}%
\providecommand \href@noop [0]{\@secondoftwo}%
\providecommand \href [0]{\begingroup \@sanitize@url \@href}%
\providecommand \@href[1]{\@@startlink{#1}\@@href}%
\providecommand \@@href[1]{\endgroup#1\@@endlink}%
\providecommand \@sanitize@url [0]{\catcode `\\12\catcode `\$12\catcode `\&12\catcode `\#12\catcode `\^12\catcode `\_12\catcode `\%12\relax}%
\providecommand \@@startlink[1]{}%
\providecommand \@@endlink[0]{}%
\providecommand \url  [0]{\begingroup\@sanitize@url \@url }%
\providecommand \@url [1]{\endgroup\@href {#1}{\urlprefix }}%
\providecommand \urlprefix  [0]{URL }%
\providecommand \Eprint [0]{\href }%
\providecommand \doibase [0]{http://dx.doi.org/}%
\providecommand \selectlanguage [0]{\@gobble}%
\providecommand \bibinfo  [0]{\@secondoftwo}%
\providecommand \bibfield  [0]{\@secondoftwo}%
\providecommand \translation [1]{[#1]}%
\providecommand \BibitemOpen [0]{}%
\providecommand \bibitemStop [0]{}%
\providecommand \bibitemNoStop [0]{.\EOS\space}%
\providecommand \EOS [0]{\spacefactor3000\relax}%
\providecommand \BibitemShut  [1]{\csname bibitem#1\endcsname}%
\let\auto@bib@innerbib\@empty
\bibitem [{\citenamefont {Edwards}\ \emph {et~al.}(2024)\citenamefont {Edwards}, \citenamefont {Fasano}, \citenamefont {Giakas}, \citenamefont {Wang}, \citenamefont {Griff-McMahon}, \citenamefont {Morozov}, \citenamefont {Perez-Ramirez}, \citenamefont {Lemos}, \citenamefont {Michel},\ and\ \citenamefont {Mikhailova}}]{Edwards24}%
  \BibitemOpen
  \bibfield  {author} {\bibinfo {author} {\bibfnamefont {M.~R.}\ \bibnamefont {Edwards}}, \bibinfo {author} {\bibfnamefont {N.~M.}\ \bibnamefont {Fasano}}, \bibinfo {author} {\bibfnamefont {A.~M.}\ \bibnamefont {Giakas}}, \bibinfo {author} {\bibfnamefont {M.~M.}\ \bibnamefont {Wang}}, \bibinfo {author} {\bibfnamefont {J.}~\bibnamefont {Griff-McMahon}}, \bibinfo {author} {\bibfnamefont {A.}~\bibnamefont {Morozov}}, \bibinfo {author} {\bibfnamefont {V.~M.}\ \bibnamefont {Perez-Ramirez}}, \bibinfo {author} {\bibfnamefont {N.}~\bibnamefont {Lemos}}, \bibinfo {author} {\bibfnamefont {P.}~\bibnamefont {Michel}}, \ and\ \bibinfo {author} {\bibfnamefont {J.~M.}\ \bibnamefont {Mikhailova}},\ }\href@noop {} {\bibfield  {journal} {\bibinfo  {journal} {Phys. Rev. Lett}\ }\textbf {\bibinfo {volume} {133}},\ \bibinfo {pages} {155101} (\bibinfo {year} {2024})}\BibitemShut {NoStop}%
\bibitem [{\citenamefont {Wu}\ \emph {et~al.}(2022)\citenamefont {Wu}, \citenamefont {Zeng}, \citenamefont {Li}, \citenamefont {Zhang}, \citenamefont {Wang}, \citenamefont {Hu}, \citenamefont {Wang}, \citenamefont {Mu}, \citenamefont {Su}, \citenamefont {Zhu}, \citenamefont {Wei}, ,\ and\ \citenamefont {Zuo}}]{Zhaohui22}%
  \BibitemOpen
  \bibfield  {author} {\bibinfo {author} {\bibfnamefont {Z.}~\bibnamefont {Wu}}, \bibinfo {author} {\bibfnamefont {X.}~\bibnamefont {Zeng}}, \bibinfo {author} {\bibfnamefont {Z.}~\bibnamefont {Li}}, \bibinfo {author} {\bibfnamefont {Z.}~\bibnamefont {Zhang}}, \bibinfo {author} {\bibfnamefont {X.}~\bibnamefont {Wang}}, \bibinfo {author} {\bibfnamefont {B.}~\bibnamefont {Hu}}, \bibinfo {author} {\bibfnamefont {X.}~\bibnamefont {Wang}}, \bibinfo {author} {\bibfnamefont {J.}~\bibnamefont {Mu}}, \bibinfo {author} {\bibfnamefont {J.}~\bibnamefont {Su}}, \bibinfo {author} {\bibfnamefont {Q.}~\bibnamefont {Zhu}}, \bibinfo {author} {\bibfnamefont {X.}~\bibnamefont {Wei}}, , \ and\ \bibinfo {author} {\bibfnamefont {Y.}~\bibnamefont {Zuo}},\ }\href@noop {} {\bibfield  {journal} {\bibinfo  {journal} {Matter Radiat. Extremes}\ }\textbf {\bibinfo {volume} {7}},\ \bibinfo {pages} {064402} (\bibinfo {year} {2022})}\BibitemShut {NoStop}%
\bibitem [{\citenamefont {Andreev}\ \emph {et~al.}(2006)\citenamefont {Andreev}, \citenamefont {Riconda}, \citenamefont {Tikhonchuk},\ and\ \citenamefont {Weber}}]{Andreev06}%
  \BibitemOpen
  \bibfield  {author} {\bibinfo {author} {\bibfnamefont {A.~A.}\ \bibnamefont {Andreev}}, \bibinfo {author} {\bibfnamefont {C.}~\bibnamefont {Riconda}}, \bibinfo {author} {\bibfnamefont {V.~T.}\ \bibnamefont {Tikhonchuk}}, \ and\ \bibinfo {author} {\bibfnamefont {S.}~\bibnamefont {Weber}},\ }\href@noop {} {\bibfield  {journal} {\bibinfo  {journal} {Phys. Plasma}\ }\textbf {\bibinfo {volume} {13}},\ \bibinfo {pages} {053110} (\bibinfo {year} {2006})}\BibitemShut {NoStop}%
\bibitem [{\citenamefont {Malkin}, \citenamefont {Shvets},\ and\ \citenamefont {Fisch}(1999)}]{Malkin991}%
  \BibitemOpen
  \bibfield  {author} {\bibinfo {author} {\bibfnamefont {V.~M.}\ \bibnamefont {Malkin}}, \bibinfo {author} {\bibfnamefont {G.}~\bibnamefont {Shvets}}, \ and\ \bibinfo {author} {\bibfnamefont {N.~J.}\ \bibnamefont {Fisch}},\ }\href@noop {} {\bibfield  {journal} {\bibinfo  {journal} {Phys. Rev. Lett}\ }\textbf {\bibinfo {volume} {82}},\ \bibinfo {pages} {4448} (\bibinfo {year} {1999})}\BibitemShut {NoStop}%
\bibitem [{\citenamefont {Ren}\ \emph {et~al.}(2007)\citenamefont {Ren}, \citenamefont {Cheng}, \citenamefont {Li},\ and\ \citenamefont {Suckewer}}]{JUN07}%
  \BibitemOpen
  \bibfield  {author} {\bibinfo {author} {\bibfnamefont {J.}~\bibnamefont {Ren}}, \bibinfo {author} {\bibfnamefont {W.}~\bibnamefont {Cheng}}, \bibinfo {author} {\bibfnamefont {S.}~\bibnamefont {Li}}, \ and\ \bibinfo {author} {\bibfnamefont {S.}~\bibnamefont {Suckewer}},\ }\href@noop {} {\bibfield  {journal} {\bibinfo  {journal} {Nat. Phys.}\ }\textbf {\bibinfo {volume} {3}},\ \bibinfo {pages} {732} (\bibinfo {year} {2007})}\BibitemShut {NoStop}%
\bibitem [{\citenamefont {Trines}\ \emph {et~al.}(2011)\citenamefont {Trines}, \citenamefont {Fiuza}, \citenamefont {Bingham}, \citenamefont {Fonseca}, \citenamefont {Silva}, \citenamefont {Cairns},\ and\ \citenamefont {Norreys}}]{Trines111}%
  \BibitemOpen
  \bibfield  {author} {\bibinfo {author} {\bibfnamefont {R.~M. G.~M.}\ \bibnamefont {Trines}}, \bibinfo {author} {\bibfnamefont {F.}~\bibnamefont {Fiuza}}, \bibinfo {author} {\bibfnamefont {R.}~\bibnamefont {Bingham}}, \bibinfo {author} {\bibfnamefont {R.~A.}\ \bibnamefont {Fonseca}}, \bibinfo {author} {\bibfnamefont {L.~O.}\ \bibnamefont {Silva}}, \bibinfo {author} {\bibfnamefont {R.~A.}\ \bibnamefont {Cairns}}, \ and\ \bibinfo {author} {\bibfnamefont {P.~A.}\ \bibnamefont {Norreys}},\ }\href@noop {} {\bibfield  {journal} {\bibinfo  {journal} {Nat. Phys.}\ }\textbf {\bibinfo {volume} {7}},\ \bibinfo {pages} {87} (\bibinfo {year} {2011})}\BibitemShut {NoStop}%
\bibitem [{\citenamefont {Toroker}, \citenamefont {Malkin},\ and\ \citenamefont {Fisch}(2014)}]{Toroker14}%
  \BibitemOpen
  \bibfield  {author} {\bibinfo {author} {\bibfnamefont {Z.}~\bibnamefont {Toroker}}, \bibinfo {author} {\bibfnamefont {V.~M.}\ \bibnamefont {Malkin}}, \ and\ \bibinfo {author} {\bibfnamefont {N.~J.}\ \bibnamefont {Fisch}},\ }\href@noop {} {\bibfield  {journal} {\bibinfo  {journal} {Phys. Plasmas}\ }\textbf {\bibinfo {volume} {21}},\ \bibinfo {pages} {113110} (\bibinfo {year} {2014})}\BibitemShut {NoStop}%
\bibitem [{\citenamefont {Edwards}\ \emph {et~al.}(2015)\citenamefont {Edwards}, \citenamefont {Toroker}, \citenamefont {Mikhailova},\ and\ \citenamefont {Fisch}}]{Matthew15}%
  \BibitemOpen
  \bibfield  {author} {\bibinfo {author} {\bibfnamefont {M.~R.}\ \bibnamefont {Edwards}}, \bibinfo {author} {\bibfnamefont {Z.}~\bibnamefont {Toroker}}, \bibinfo {author} {\bibfnamefont {J.~M.}\ \bibnamefont {Mikhailova}}, \ and\ \bibinfo {author} {\bibfnamefont {N.~J.}\ \bibnamefont {Fisch}},\ }\href@noop {} {\bibfield  {journal} {\bibinfo  {journal} {Phys. Plasmas}\ }\textbf {\bibinfo {volume} {22}},\ \bibinfo {pages} {074501} (\bibinfo {year} {2015})}\BibitemShut {NoStop}%
\bibitem [{\citenamefont {Marqu\'{e}s}\ \emph {et~al.}(2019)\citenamefont {Marqu\'{e}s}, \citenamefont {Lancia}, \citenamefont {Gangolf}, \citenamefont {Blecher}, \citenamefont {Bolanos}, \citenamefont {J.Fuchs}, \citenamefont {Willi}, \citenamefont {Amiranoff}, \citenamefont {Berger}, \citenamefont {Chiaramello}, \citenamefont {Weber},\ and\ \citenamefont {Riconda}}]{Marques19}%
  \BibitemOpen
  \bibfield  {author} {\bibinfo {author} {\bibfnamefont {J.-R.}\ \bibnamefont {Marqu\'{e}s}}, \bibinfo {author} {\bibfnamefont {L.}~\bibnamefont {Lancia}}, \bibinfo {author} {\bibfnamefont {T.}~\bibnamefont {Gangolf}}, \bibinfo {author} {\bibfnamefont {M.}~\bibnamefont {Blecher}}, \bibinfo {author} {\bibfnamefont {S.}~\bibnamefont {Bolanos}}, \bibinfo {author} {\bibnamefont {J.Fuchs}}, \bibinfo {author} {\bibfnamefont {O.}~\bibnamefont {Willi}}, \bibinfo {author} {\bibfnamefont {F.}~\bibnamefont {Amiranoff}}, \bibinfo {author} {\bibfnamefont {R.~L.}\ \bibnamefont {Berger}}, \bibinfo {author} {\bibfnamefont {M.}~\bibnamefont {Chiaramello}}, \bibinfo {author} {\bibfnamefont {S.}~\bibnamefont {Weber}}, \ and\ \bibinfo {author} {\bibfnamefont {C.}~\bibnamefont {Riconda}},\ }\href@noop {} {\bibfield  {journal} {\bibinfo  {journal} {Phys. Rev. X}\ }\textbf {\bibinfo {volume} {9}},\ \bibinfo {pages} {021008} (\bibinfo {year} {2019})}\BibitemShut {NoStop}%
\bibitem [{\citenamefont {Wu}\ \emph {et~al.}(2024)\citenamefont {Wu}, \citenamefont {Peng}, \citenamefont {Zeng}, \citenamefont {Li}, \citenamefont {Zhang}, \citenamefont {Wang}, \citenamefont {Wang}, \citenamefont {Mu}, \citenamefont {Zuo}, \citenamefont {Su}, \citenamefont {Cao}, \citenamefont {Fu}, \citenamefont {Riconda},\ and\ \citenamefont {Weber}}]{Wu24PRR}%
  \BibitemOpen
  \bibfield  {author} {\bibinfo {author} {\bibfnamefont {Z.}~\bibnamefont {Wu}}, \bibinfo {author} {\bibfnamefont {H.}~\bibnamefont {Peng}}, \bibinfo {author} {\bibfnamefont {X.}~\bibnamefont {Zeng}}, \bibinfo {author} {\bibfnamefont {Z.}~\bibnamefont {Li}}, \bibinfo {author} {\bibfnamefont {Z.}~\bibnamefont {Zhang}}, \bibinfo {author} {\bibfnamefont {X.}~\bibnamefont {Wang}}, \bibinfo {author} {\bibfnamefont {X.}~\bibnamefont {Wang}}, \bibinfo {author} {\bibfnamefont {J.}~\bibnamefont {Mu}}, \bibinfo {author} {\bibfnamefont {Y.}~\bibnamefont {Zuo}}, \bibinfo {author} {\bibfnamefont {J.}~\bibnamefont {Su}}, \bibinfo {author} {\bibfnamefont {H.}~\bibnamefont {Cao}}, \bibinfo {author} {\bibfnamefont {Y.}~\bibnamefont {Fu}}, \bibinfo {author} {\bibfnamefont {C.}~\bibnamefont {Riconda}}, \ and\ \bibinfo {author} {\bibfnamefont {S.}~\bibnamefont {Weber}},\ }\href@noop {} {\bibfield  {journal} {\bibinfo  {journal} {Phys. Rev. Research}\ }\textbf {\bibinfo {volume} {6}},\ \bibinfo {pages} {013126} (\bibinfo {year}
  {2024})}\BibitemShut {NoStop}%
\bibitem [{\citenamefont {Turnbull}\ \emph {et~al.}(2016)\citenamefont {Turnbull}, \citenamefont {Michel}, \citenamefont {Chapman}, \citenamefont {Tubman}, \citenamefont {Pollock},\ and\ \citenamefont {Chen}}]{Turnbull16}%
  \BibitemOpen
  \bibfield  {author} {\bibinfo {author} {\bibfnamefont {D.}~\bibnamefont {Turnbull}}, \bibinfo {author} {\bibfnamefont {P.}~\bibnamefont {Michel}}, \bibinfo {author} {\bibfnamefont {T.}~\bibnamefont {Chapman}}, \bibinfo {author} {\bibfnamefont {E.}~\bibnamefont {Tubman}}, \bibinfo {author} {\bibfnamefont {B.~B.}\ \bibnamefont {Pollock}}, \ and\ \bibinfo {author} {\bibfnamefont {C.~Y.}\ \bibnamefont {Chen}},\ }\href@noop {} {\bibfield  {journal} {\bibinfo  {journal} {Phys. Rev. Lett}\ }\textbf {\bibinfo {volume} {116}},\ \bibinfo {pages} {205001} (\bibinfo {year} {2016})}\BibitemShut {NoStop}%
\bibitem [{\citenamefont {Turnbull}\ \emph {et~al.}(2017)\citenamefont {Turnbull}, \citenamefont {Goyon}, \citenamefont {Kemp}, \citenamefont {Pollock}, \citenamefont {Mariscal}, \citenamefont {Divol}, \citenamefont {Ross}, \citenamefont {Patankar}, \citenamefont {Moody},\ and\ \citenamefont {Michel}}]{Turnbull17}%
  \BibitemOpen
  \bibfield  {author} {\bibinfo {author} {\bibfnamefont {D.}~\bibnamefont {Turnbull}}, \bibinfo {author} {\bibfnamefont {C.}~\bibnamefont {Goyon}}, \bibinfo {author} {\bibfnamefont {G.~E.}\ \bibnamefont {Kemp}}, \bibinfo {author} {\bibfnamefont {B.~B.}\ \bibnamefont {Pollock}}, \bibinfo {author} {\bibfnamefont {D.}~\bibnamefont {Mariscal}}, \bibinfo {author} {\bibfnamefont {L.}~\bibnamefont {Divol}}, \bibinfo {author} {\bibfnamefont {J.~S.}\ \bibnamefont {Ross}}, \bibinfo {author} {\bibfnamefont {S.}~\bibnamefont {Patankar}}, \bibinfo {author} {\bibfnamefont {J.~D.}\ \bibnamefont {Moody}}, \ and\ \bibinfo {author} {\bibfnamefont {P.}~\bibnamefont {Michel}},\ }\href@noop {} {\bibfield  {journal} {\bibinfo  {journal} {Phys. Rev. Lett}\ }\textbf {\bibinfo {volume} {118}},\ \bibinfo {pages} {015001} (\bibinfo {year} {2017})}\BibitemShut {NoStop}%
\bibitem [{\citenamefont {Lehmann}\ and\ \citenamefont {Spatschek}(2018)}]{Lehmann18}%
  \BibitemOpen
  \bibfield  {author} {\bibinfo {author} {\bibfnamefont {G.}~\bibnamefont {Lehmann}}\ and\ \bibinfo {author} {\bibfnamefont {K.~H.}\ \bibnamefont {Spatschek}},\ }\href@noop {} {\bibfield  {journal} {\bibinfo  {journal} {Phys. Rev. E}\ }\textbf {\bibinfo {volume} {97}},\ \bibinfo {pages} {063201} (\bibinfo {year} {2018})}\BibitemShut {NoStop}%
\bibitem [{\citenamefont {Liu}\ \emph {et~al.}(2010)\citenamefont {Liu}, \citenamefont {Durand}, \citenamefont {Chen}, \citenamefont {Houard}, \citenamefont {Prade}, \citenamefont {Forestier},\ and\ \citenamefont {Mysyrowicz}}]{Liu10}%
  \BibitemOpen
  \bibfield  {author} {\bibinfo {author} {\bibfnamefont {Y.}~\bibnamefont {Liu}}, \bibinfo {author} {\bibfnamefont {M.}~\bibnamefont {Durand}}, \bibinfo {author} {\bibfnamefont {S.}~\bibnamefont {Chen}}, \bibinfo {author} {\bibfnamefont {A.}~\bibnamefont {Houard}}, \bibinfo {author} {\bibfnamefont {B.}~\bibnamefont {Prade}}, \bibinfo {author} {\bibfnamefont {B.}~\bibnamefont {Forestier}}, \ and\ \bibinfo {author} {\bibfnamefont {A.}~\bibnamefont {Mysyrowicz}},\ }\href@noop {} {\bibfield  {journal} {\bibinfo  {journal} {Phys. Rev. Lett}\ }\textbf {\bibinfo {volume} {105}},\ \bibinfo {pages} {055003} (\bibinfo {year} {2010})}\BibitemShut {NoStop}%
\bibitem [{\citenamefont {Vincenti}(2019)}]{Henri19}%
  \BibitemOpen
  \bibfield  {author} {\bibinfo {author} {\bibfnamefont {H.}~\bibnamefont {Vincenti}},\ }\href@noop {} {\bibfield  {journal} {\bibinfo  {journal} {Phys. Rev. Lett}\ }\textbf {\bibinfo {volume} {123}},\ \bibinfo {pages} {105001} (\bibinfo {year} {2019})}\BibitemShut {NoStop}%
\bibitem [{\citenamefont {Edwards}\ \emph {et~al.}(2022)\citenamefont {Edwards}, \citenamefont {Munirov}, \citenamefont {Singh}, \citenamefont {Fasano}, \citenamefont {Kur}, \citenamefont {Lemos}, \citenamefont {Mikhailova}, \citenamefont {Wurtele},\ and\ \citenamefont {Michel}}]{Edwards22}%
  \BibitemOpen
  \bibfield  {author} {\bibinfo {author} {\bibfnamefont {M.~R.}\ \bibnamefont {Edwards}}, \bibinfo {author} {\bibfnamefont {V.~R.}\ \bibnamefont {Munirov}}, \bibinfo {author} {\bibfnamefont {A.}~\bibnamefont {Singh}}, \bibinfo {author} {\bibfnamefont {N.~M.}\ \bibnamefont {Fasano}}, \bibinfo {author} {\bibfnamefont {E.}~\bibnamefont {Kur}}, \bibinfo {author} {\bibfnamefont {N.}~\bibnamefont {Lemos}}, \bibinfo {author} {\bibfnamefont {J.~M.}\ \bibnamefont {Mikhailova}}, \bibinfo {author} {\bibfnamefont {J.~S.}\ \bibnamefont {Wurtele}}, \ and\ \bibinfo {author} {\bibfnamefont {P.}~\bibnamefont {Michel}},\ }\href@noop {} {\bibfield  {journal} {\bibinfo  {journal} {Phys. Rev. Lett}\ }\textbf {\bibinfo {volume} {128}},\ \bibinfo {pages} {065003} (\bibinfo {year} {2022})}\BibitemShut {NoStop}%
\bibitem [{\citenamefont {Qu}\ and\ \citenamefont {Fisch}(2024)}]{qu2024entangled}%
  \BibitemOpen
  \bibfield  {author} {\bibinfo {author} {\bibfnamefont {K.}~\bibnamefont {Qu}}\ and\ \bibinfo {author} {\bibfnamefont {N.~J.}\ \bibnamefont {Fisch}},\ }\href {\doibase 10.1103/PhysRevE.110.065211} {\bibfield  {journal} {\bibinfo  {journal} {Physical Review E}\ }\textbf {\bibinfo {volume} {110}},\ \bibinfo {pages} {065211} (\bibinfo {year} {2024})}\BibitemShut {NoStop}%
\bibitem [{\citenamefont {Leblanc}\ \emph {et~al.}(2017)\citenamefont {Leblanc}, \citenamefont {Denoeud}, \citenamefont {Chopineau}, \citenamefont {Mennerat}, \citenamefont {Martin},\ and\ \citenamefont {Qu\'{e}r\'{e}}}]{Leblanc17}%
  \BibitemOpen
  \bibfield  {author} {\bibinfo {author} {\bibfnamefont {A.}~\bibnamefont {Leblanc}}, \bibinfo {author} {\bibfnamefont {A.}~\bibnamefont {Denoeud}}, \bibinfo {author} {\bibfnamefont {L.}~\bibnamefont {Chopineau}}, \bibinfo {author} {\bibfnamefont {G.}~\bibnamefont {Mennerat}}, \bibinfo {author} {\bibfnamefont {P.}~\bibnamefont {Martin}}, \ and\ \bibinfo {author} {\bibfnamefont {F.}~\bibnamefont {Qu\'{e}r\'{e}}},\ }\href@noop {} {\bibfield  {journal} {\bibinfo  {journal} {Nature physics}\ }\textbf {\bibinfo {volume} {13}},\ \bibinfo {pages} {440} (\bibinfo {year} {2017})}\BibitemShut {NoStop}%
\bibitem [{\citenamefont {Dodin}\ and\ \citenamefont {Fisch}(2002{\natexlab{a}})}]{Dodin02prl}%
  \BibitemOpen
  \bibfield  {author} {\bibinfo {author} {\bibfnamefont {I.~Y.}\ \bibnamefont {Dodin}}\ and\ \bibinfo {author} {\bibfnamefont {N.~J.}\ \bibnamefont {Fisch}},\ }\href@noop {} {\bibfield  {journal} {\bibinfo  {journal} {Phys. Rev. Lett.}\ }\textbf {\bibinfo {volume} {88}},\ \bibinfo {pages} {165001} (\bibinfo {year} {2002}{\natexlab{a}})}\BibitemShut {NoStop}%
\bibitem [{\citenamefont {Dodin}\ and\ \citenamefont {Fisch}(2002{\natexlab{b}})}]{dodin2002dynamic}%
  \BibitemOpen
  \bibfield  {author} {\bibinfo {author} {\bibfnamefont {I.~Y.}\ \bibnamefont {Dodin}}\ and\ \bibinfo {author} {\bibfnamefont {N.~J.}\ \bibnamefont {Fisch}},\ }\href {\doibase 10.1016/S0030-4018(02)02144-2} {\bibfield  {journal} {\bibinfo  {journal} {Optics Communications}\ }\textbf {\bibinfo {volume} {214}},\ \bibinfo {pages} {83} (\bibinfo {year} {2002}{\natexlab{b}})}\BibitemShut {NoStop}%
\bibitem [{\citenamefont {Lehmann}\ and\ \citenamefont {Spatschek}(2019)}]{Lehmann19}%
  \BibitemOpen
  \bibfield  {author} {\bibinfo {author} {\bibfnamefont {G.}~\bibnamefont {Lehmann}}\ and\ \bibinfo {author} {\bibfnamefont {K.~H.}\ \bibnamefont {Spatschek}},\ }\href@noop {} {\bibfield  {journal} {\bibinfo  {journal} {Phys. Rev. E}\ }\textbf {\bibinfo {volume} {100}},\ \bibinfo {pages} {033205} (\bibinfo {year} {2019})}\BibitemShut {NoStop}%
\bibitem [{\citenamefont {Shi}\ \emph {et~al.}(2011)\citenamefont {Shi}, \citenamefont {Li}, \citenamefont {Wang}, \citenamefont {Lu}, \citenamefont {Ding},\ and\ \citenamefont {Zeng}}]{Liping11}%
  \BibitemOpen
  \bibfield  {author} {\bibinfo {author} {\bibfnamefont {L.}~\bibnamefont {Shi}}, \bibinfo {author} {\bibfnamefont {W.}~\bibnamefont {Li}}, \bibinfo {author} {\bibfnamefont {Y.}~\bibnamefont {Wang}}, \bibinfo {author} {\bibfnamefont {X.}~\bibnamefont {Lu}}, \bibinfo {author} {\bibfnamefont {L.}~\bibnamefont {Ding}}, \ and\ \bibinfo {author} {\bibfnamefont {H.}~\bibnamefont {Zeng}},\ }\href@noop {} {\bibfield  {journal} {\bibinfo  {journal} {Optica}\ }\textbf {\bibinfo {volume} {107}},\ \bibinfo {pages} {095004} (\bibinfo {year} {2011})}\BibitemShut {NoStop}%
\bibitem [{\citenamefont {Zhang}\ \emph {et~al.}(2021)\citenamefont {Zhang}, \citenamefont {Nie}, \citenamefont {Wu}, \citenamefont {Sinclair}, \citenamefont {Huang}, \citenamefont {Marsh},\ and\ \citenamefont {Joshi}}]{Chaojie21}%
  \BibitemOpen
  \bibfield  {author} {\bibinfo {author} {\bibfnamefont {C.}~\bibnamefont {Zhang}}, \bibinfo {author} {\bibfnamefont {Z.}~\bibnamefont {Nie}}, \bibinfo {author} {\bibfnamefont {Y.}~\bibnamefont {Wu}}, \bibinfo {author} {\bibfnamefont {M.}~\bibnamefont {Sinclair}}, \bibinfo {author} {\bibfnamefont {C.-K.}\ \bibnamefont {Huang}}, \bibinfo {author} {\bibfnamefont {K.~A.}\ \bibnamefont {Marsh}}, \ and\ \bibinfo {author} {\bibfnamefont {C.}~\bibnamefont {Joshi}},\ }\href@noop {} {\bibfield  {journal} {\bibinfo  {journal} {Plasma Phys. Control. Fusion}\ }\textbf {\bibinfo {volume} {63}},\ \bibinfo {pages} {095011} (\bibinfo {year} {2021})}\BibitemShut {NoStop}%
\bibitem [{\citenamefont {Edwards}\ \emph {et~al.}(2023)\citenamefont {Edwards}, \citenamefont {Waczynski}, \citenamefont {Rockafellow}, \citenamefont {Manzo}, \citenamefont {Zingale}, \citenamefont {Michel},\ and\ \citenamefont {Milchberg}}]{Edwards23}%
  \BibitemOpen
  \bibfield  {author} {\bibinfo {author} {\bibfnamefont {M.~R.}\ \bibnamefont {Edwards}}, \bibinfo {author} {\bibfnamefont {S.}~\bibnamefont {Waczynski}}, \bibinfo {author} {\bibfnamefont {E.}~\bibnamefont {Rockafellow}}, \bibinfo {author} {\bibfnamefont {L.}~\bibnamefont {Manzo}}, \bibinfo {author} {\bibfnamefont {A.}~\bibnamefont {Zingale}}, \bibinfo {author} {\bibfnamefont {P.}~\bibnamefont {Michel}}, \ and\ \bibinfo {author} {\bibfnamefont {H.~M.}\ \bibnamefont {Milchberg}},\ }\href@noop {} {\bibfield  {journal} {\bibinfo  {journal} {Optica}\ }\textbf {\bibinfo {volume} {10}},\ \bibinfo {pages} {1587} (\bibinfo {year} {2023})}\BibitemShut {NoStop}%
\bibitem [{\citenamefont {Clark}\ and\ \citenamefont {Fisch}(2002)}]{clark02}%
  \BibitemOpen
  \bibfield  {author} {\bibinfo {author} {\bibfnamefont {D.~S.}\ \bibnamefont {Clark}}\ and\ \bibinfo {author} {\bibfnamefont {N.~J.}\ \bibnamefont {Fisch}},\ }\href@noop {} {\bibfield  {journal} {\bibinfo  {journal} {Phys. Plasmas}\ }\textbf {\bibinfo {volume} {9}},\ \bibinfo {pages} {2772} (\bibinfo {year} {2002})}\BibitemShut {NoStop}%
\bibitem [{\citenamefont {Clark}(2003)}]{Clark032}%
  \BibitemOpen
  \bibfield  {author} {\bibinfo {author} {\bibfnamefont {D.~S.}\ \bibnamefont {Clark}},\ }\emph {\bibinfo {title} {Investigations of Raman Laser Amplification in Preformed and Ionizing Plasmas}},\ \href@noop {} {Ph.D. thesis},\ \bibinfo  {school} {Princeton University} (\bibinfo {year} {2003})\BibitemShut {NoStop}%
\bibitem [{\citenamefont {Zhao}, \citenamefont {Liu},\ and\ \citenamefont {Zhou}(2014)}]{zhao2014multiphoton}%
  \BibitemOpen
  \bibfield  {author} {\bibinfo {author} {\bibfnamefont {S.-F.}\ \bibnamefont {Zhao}}, \bibinfo {author} {\bibfnamefont {L.}~\bibnamefont {Liu}}, \ and\ \bibinfo {author} {\bibfnamefont {X.-X.}\ \bibnamefont {Zhou}},\ }\href {\doibase 10.1016/j.optcom.2013.09.074} {\bibfield  {journal} {\bibinfo  {journal} {Optics Communications}\ }\textbf {\bibinfo {volume} {313}},\ \bibinfo {pages} {74} (\bibinfo {year} {2014})}\BibitemShut {NoStop}%
\bibitem [{\citenamefont {Zhao}\ \emph {et~al.}(2016{\natexlab{a}})\citenamefont {Zhao}, \citenamefont {Le}, \citenamefont {Jin}, \citenamefont {Wang},\ and\ \citenamefont {Lin}}]{zhao2016analytical}%
  \BibitemOpen
  \bibfield  {author} {\bibinfo {author} {\bibfnamefont {S.-F.}\ \bibnamefont {Zhao}}, \bibinfo {author} {\bibfnamefont {A.-T.}\ \bibnamefont {Le}}, \bibinfo {author} {\bibfnamefont {C.}~\bibnamefont {Jin}}, \bibinfo {author} {\bibfnamefont {X.}~\bibnamefont {Wang}}, \ and\ \bibinfo {author} {\bibfnamefont {C.~D.}\ \bibnamefont {Lin}},\ }\href {\doibase 10.1103/PhysRevA.93.023413} {\bibfield  {journal} {\bibinfo  {journal} {Physical Review A}\ }\textbf {\bibinfo {volume} {93}},\ \bibinfo {pages} {023413} (\bibinfo {year} {2016}{\natexlab{a}})}\BibitemShut {NoStop}%
\bibitem [{\citenamefont {Zhao}\ \emph {et~al.}(2016{\natexlab{b}})\citenamefont {Zhao}, \citenamefont {Le}, \citenamefont {Jin}, \citenamefont {Wang},\ and\ \citenamefont {Lin}}]{zhao2016kinetics}%
  \BibitemOpen
  \bibfield  {author} {\bibinfo {author} {\bibfnamefont {S.-F.}\ \bibnamefont {Zhao}}, \bibinfo {author} {\bibfnamefont {A.-T.}\ \bibnamefont {Le}}, \bibinfo {author} {\bibfnamefont {C.}~\bibnamefont {Jin}}, \bibinfo {author} {\bibfnamefont {X.}~\bibnamefont {Wang}}, \ and\ \bibinfo {author} {\bibfnamefont {C.~D.}\ \bibnamefont {Lin}},\ }\href {\doibase 10.1063/1.4953475} {\bibfield  {journal} {\bibinfo  {journal} {Journal of Applied Physics}\ }\textbf {\bibinfo {volume} {125}},\ \bibinfo {pages} {243301} (\bibinfo {year} {2016}{\natexlab{b}})}\BibitemShut {NoStop}%
\bibitem [{\citenamefont {Popov}(2004)}]{popov2004tunnel}%
  \BibitemOpen
  \bibfield  {author} {\bibinfo {author} {\bibfnamefont {V.~S.}\ \bibnamefont {Popov}},\ }\href {\doibase 10.1070/PU2004v047n09ABEH001789} {\bibfield  {journal} {\bibinfo  {journal} {Physics-Uspekhi}\ }\textbf {\bibinfo {volume} {47}},\ \bibinfo {pages} {855} (\bibinfo {year} {2004})}\BibitemShut {NoStop}%
\bibitem [{\citenamefont {Tong}\ and\ \citenamefont {Lin}(2005)}]{tong2005empirical}%
  \BibitemOpen
  \bibfield  {author} {\bibinfo {author} {\bibfnamefont {X.~M.}\ \bibnamefont {Tong}}\ and\ \bibinfo {author} {\bibfnamefont {C.~D.}\ \bibnamefont {Lin}},\ }\href {\doibase 10.1088/0953-4075/38/15/006} {\bibfield  {journal} {\bibinfo  {journal} {Journal of Physics B: Atomic, Molecular and Optical Physics}\ }\textbf {\bibinfo {volume} {38}},\ \bibinfo {pages} {2593} (\bibinfo {year} {2005})}\BibitemShut {NoStop}%
\bibitem [{\citenamefont {Zhou}\ \emph {et~al.}(2009)\citenamefont {Zhou}, \citenamefont {Lock}, \citenamefont {Li}, \citenamefont {Wagner}, \citenamefont {Murnane},\ and\ \citenamefont {Kapteyn}}]{zhou2009molecular}%
  \BibitemOpen
  \bibfield  {author} {\bibinfo {author} {\bibfnamefont {Y.}~\bibnamefont {Zhou}}, \bibinfo {author} {\bibfnamefont {R.}~\bibnamefont {Lock}}, \bibinfo {author} {\bibfnamefont {W.}~\bibnamefont {Li}}, \bibinfo {author} {\bibfnamefont {N.}~\bibnamefont {Wagner}}, \bibinfo {author} {\bibfnamefont {M.~M.}\ \bibnamefont {Murnane}}, \ and\ \bibinfo {author} {\bibfnamefont {H.~C.}\ \bibnamefont {Kapteyn}},\ }\href {\doibase 10.1103/PhysRevLett.102.073902} {\bibfield  {journal} {\bibinfo  {journal} {Physical Review Letters}\ }\textbf {\bibinfo {volume} {102}},\ \bibinfo {pages} {073902} (\bibinfo {year} {2009})}\BibitemShut {NoStop}%
\end{thebibliography}%

\end{document}